\documentclass{iopjournal}
\usepackage{amsmath}
\usepackage{amssymb}
\usepackage{mathtools}
\usepackage{bm}
\usepackage{array}
\usepackage{tabularx}
\usepackage{dcolumn}
\usepackage{makecell}
\usepackage{multirow}
\usepackage{booktabs}
\newcolumntype{L}[1]{%
    >{\raggedright\arraybackslash}p{#1}%
}
\newcolumntype{Y}{%
    >{\raggedright\arraybackslash}X%
}
\usepackage{siunitx}
\usepackage[numbers,sort&compress]{natbib}
\usepackage{etoolbox}
\usepackage{xcolor}
\usepackage{etoolbox}

\makeatletter
\AtBeginDocument{%
  \patchcmd{\@outputpage}
    {{\@thehead}}
    {{\color{black}\@thehead}}
    {}{}%
}
\makeatother

\begin{document}

\articletype{Paper}
\title{Loss Mechanisms in Cryogenic Microwave Epitaxial AlN Resonators}

\author{%
Hemant Gulupalli$^1$, 
Navnil Choudhury$^1$, 
Jiacheng Xie$^2$, 
Yufeng Wu$^2$, 
Huili Grace Xing$^{3,4,5}$, 
Hong X. Tang$^2$, 
Debdeep Jena$^{3,4,5,6}$, 
Kanad Basu$^1$ and 
Wenwen Zhao$^{1,*}$%
}

\affil{$^1$Department of Electrical, Computer, and Systems Engineering, 
Rensselaer Polytechnic Institute, Troy, New York 12180, USA}

\affil{$^2$Department of Electrical and Computer Engineering, 
Yale University, New Haven, Connecticut 06520, USA}

\affil{$^3$School of Electrical and Computer Engineering, 
Cornell University, Ithaca, New York 14853, USA}

\affil{$^4$Department of Materials Science and Engineering, 
Cornell University, Ithaca, New York 14853, USA}

\affil{$^5$Kavli Institute at Cornell for Nanoscale Science, 
Cornell University, Ithaca, New York 14853, USA}

\affil{$^6$School of Applied and Engineering Physics, 
Cornell University, Ithaca, New York 14853, USA}

\affil{$^*$Author to whom correspondence should be addressed.}

\email{guluph@rpi.edu, zhaow9@rpi.edu}


\begin{abstract}
Epitaxial aluminum nitride (AlN) thin-film bulk acoustic resonators (FBARs) enable low-loss filtering for future 6G systems. They also provide a compact approach for qubit sensing at cryogenic temperatures. However, these devices are rarely characterized systematically from room temperature to cryogenic temperatures, and the mechanisms that limit their cryogenic performance remain unclear. In this work, we study a \SI{15.6}{\giga\hertz} epitaxial AlN FBAR from room temperature to cryogenic temperatures to identify losses from the AlN film and those introduced by the electrodes, anchors, and other device layers. {Small-signal RF measurements from \SI{294}{\kelvin} down to \SI{6.5}{\kelvin} show an increase in the unloaded $Q_{\max}$ from $245$ to $1729$}. A temperature-dependent model that includes phonon–phonon scattering, thermoelastic damping, dielectric loss, electrical loss, and anchor loss helps explain the measured $Q(T)$ trend and identifies a transition from the Landau–Rumer to the Akhiezer regime near \SI{270}{\kelvin}. {The model indicates that acoustic energy leakage through the anchors and electrical loss set upper bounds on $Q$. Energy transfer from the main resonant mode to spurious acoustic modes can also lower the measured $Q$. Improving anchor isolation, suppressing spurious modes, and reducing electrode loss, including through superconducting electrodes, increase the cryogenic $Q$. These results guide the design of higher-$Q$ acoustic resonators for high-frequency 6G filters and cryogenic quantum-microwave circuits.}
\end{abstract}

\section{Introduction}
Sixth-generation (6G) wireless systems are expected to move into higher mid-band frequencies (often called FR3, about $7$--$24~\si{GHz}$) to support more data traffic while still keeping good coverage, and to ease crowding in bands below $7~\si{GHz}$~\cite{Izhar2025P3FAlScN, Cui2025DySPAN_FR3}. In this regime, RF front ends face a bottleneck, as carrier frequency increases and spectrum becomes more densely packed, channel selectivity and out-of-band rejection must improve without incurring insertion loss~\cite{Barrera2025_EDL_P3FLN}. Bulk acoustic wave (BAW) technologies, particularly thin-film resonators such as Aluminum Nitride (AlN) thin-film bulk acoustic resonators (FBARs), are attractive for high-frequency RF filtering. The operating frequency scales inversely with thickness, and the thickness mode is relatively easier to confine within the device. These resonators exhibit fewer spurious modes~\cite{Loebl2001MaterialsForBAWResonatorsFilters,Rosen2005SuppressSpuriousLateralFBAR} and can handle higher power without breaking down~\cite{Ding2025APR_SHFfilters}. Achieving 6G-class filter responses above $10~\si{GHz}$ requires both strong coupling $(k_t^2)$ and high $Q$ to reduce band-pass loss and sharpen selectivity~\cite{liu2024boosting,Hakim2024Ferroelectric}. 

In acoustic filters, the quality factor ($Q$) measures stored energy relative to dissipated energy per cycle~\cite{tabrizian2009phonon}. In ladder and lattice filters, a higher internal $Q$ reduces insertion loss and improves out-of-band rejection and spectral purity, provided that the electromechanical coupling is sufficiently strong~\cite{Devitt2026SpinWaveBPF6G}. At high frequencies, AlN BAW performance is limited by both intrinsic and extrinsic losses. Intrinsic loss arises from phonon interactions and thermodynamic processes within the material and depends on temperature, frequency, and the material's thermophysical properties~\cite{Chandorkar2008MEMS_Qlimits,Liu2005LossStudy}. Extrinsic loss is associated with device structures and their interaction with the environment, particularly the anchors and electrodes~\cite{Kazemi2023AnchorLossReflectors,Peczalski2015QualityFactor}. Thus, $Q_{\max}$, the highest $Q$ extracted near resonance, reflects the combined effects of material and device losses rather than an intrinsic material property~\cite{Ruby2008ExtractingUnloadedQAcrossTech}. Reducing extrinsic losses can increase $Q_{\max}$ without changing the intrinsic material loss~\cite{Gokhale2017PnCTethersIntrinsicQLimit}. At cryogenic temperatures, phonon dissipation is reduced, so losses from anchors, electrodes, and other device structures become more important.

\begin{figure}[t]
    \centering
    \includegraphics[width=\textwidth]{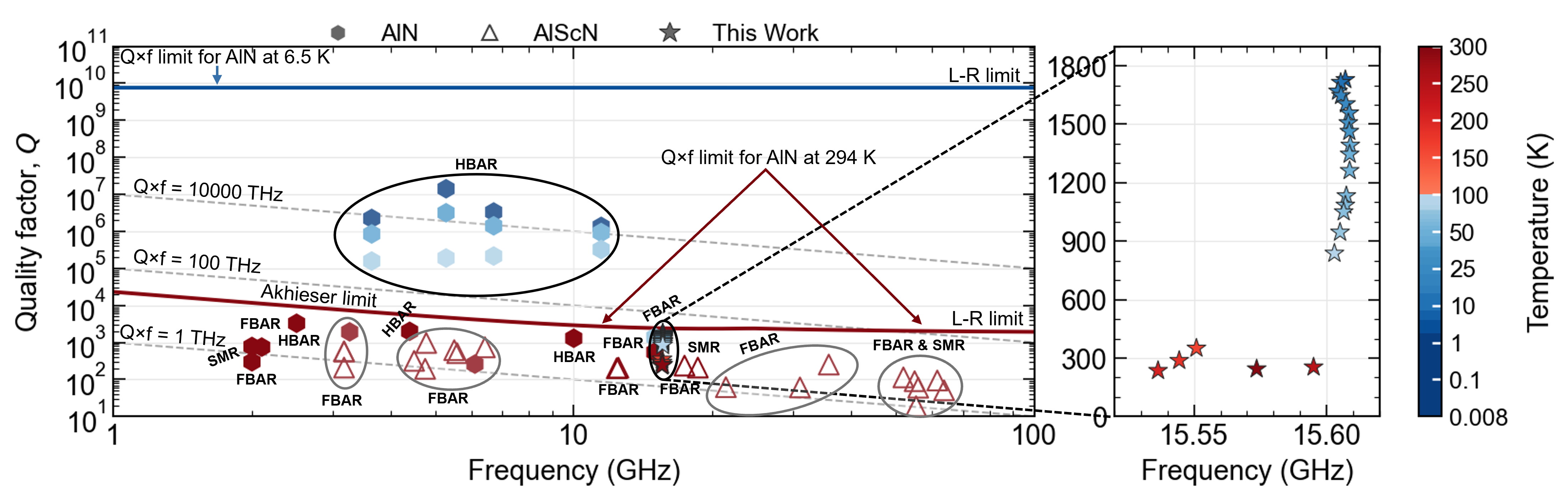}
    \caption{{Benchmarking AlN-based bulk acoustic resonator (BAW) $Q$ versus frequency at different temperatures. The highest $Q$ values reported for literature BAW devices are compared with the unloaded $Q_{\max}$ of our \SI{15.6}{\giga\hertz} AlN FBAR ($\bigstar$).}}
    \label{fig:benchmarking}
\end{figure}

{
Reducing these device losses is also important for quantum acoustic applications, where longer acoustic lifetimes allow more time for coherent interactions with superconducting qubits. Coupling piezoelectric acoustic resonators to these qubits enables the control of individual phonons~\cite{OConnell2010QuantumGroundStatePhononControl,Chu2017QuantumAcousticsSCQubits} and provides a route to hybrid quantum systems~\cite{Kervinen2018,Franse2024HighCoherenceQuantumAcousticsPlanarQubits}. Higher acoustic frequencies offer an additional thermal advantage: at a given temperature, they reduce the mean thermal phonon occupation and increase the ground-state probability~\cite{Aspelmeyer2014}, relaxing the cooling requirement for thermal initialization. Strong coupling between bulk acoustic modes near \SI{10}{\giga\hertz} and a superconducting microwave resonator has already been demonstrated~\cite{Han2016}, motivating further investigation of high-frequency acoustic interfaces with superconducting circuits.

However, cryogenic measurements of thin-film FBARs above \SI{10}{\giga\hertz} remain scarce. Here, we fabricate a \SI{15.6}{\giga\hertz} epitaxial AlN FBAR and characterize it from \SI{294}{\kelvin} to \SI{6.5}{\kelvin} to investigate the mechanisms that limit its quality factor. We combine small-signal RF measurements with a physics-based loss model that includes Landau--Rumer and Akhiezer phonon--phonon scattering, thermoelastic damping, dielectric loss, electrical loss, and anchor loss. The model uses temperature-dependent material properties and device geometry to estimate the individual loss contributions and calculate $Q_{\mathrm{total}}(T)$. Comparison with the measured $Q$ data helps explain the overall temperature dependence and which loss mechanisms dominate over the measured range. These results identify the losses to improve for resonator quality factor $Q$.
}

\section{Benchmarking Against the State of the Art}

{Figure~\ref{fig:benchmarking} benchmarks the fabricated \SI{15.6}{GHz} epitaxial AlN FBAR against AlN- and AlScN-based bulk acoustic wave (BAW) devices, comprising film bulk acoustic resonators (FBARs), solidly mounted resonators (SMRs), and high-overtone bulk acoustic resonators (HBARs). Gray dashed lines denote constant \(Qf\), while the blue and red curves provide calculated intrinsic AlN references for a longitudinal thickness mode at 6.5~K and 294~K. Unloaded quality factor (\(Q_{max}\)) values of our FBAR are included.}

{
At room temperature, AlScN FBARs achieve \(Qf\) products of a few THz across a wide frequency range. Devices operating at 4.78--6.41~GHz exhibit \(Qf\approx2.9\)--4.7~THz~\cite{Yang2026}, while devices at 12.5~GHz and 35.8~GHz reach approximately 2.6~THz and 9.0~THz, respectively~\cite{Cho2025ScAlN,Peng2025TrilayerFBAR}. Similar values are reported for solidly mounted devices: an AlScN SMR achieves \(Qf\approx3.9~\mathrm{THz}\) at 18.64~GHz~\cite{Barrera2024SMR}, while an AlN-based solidly mounted overmoded resonator reaches approximately 5.0~THz at 55~GHz~\cite{Schaffer2023OBAR}.

AlN-transduced HBARs extend this comparison to composite bulk cavities. At room temperature, devices exhibit \(Qf\approx8.3\)--13~THz over 2.5--10~GHz~\cite{Reinhardt2013HBAR,Kurosu2023ImpedanceMatchedHBAR}. Much larger values are reported at cryogenic temperatures: at approximately 10~mK, AlN/sapphire HBARs reach internal \(Qf\approx8\times10^{15}\)--\(7.5\times10^{16}\)~Hz over 3.64--11.47~GHz~\cite{GarciaBelles2026}. Other acoustic resonators, such as AlN contour-mode resonators and AlScN Lamb-wave resonators, exhibit room-temperature \(Qf\) products of order \(10^{12}\)--\(10^{13}\)~Hz~\cite{5361526,Giribaldi2024Filters}, while cryogenic AlN-based surface acoustic wave resonators can exceed \(10^{14}\)~Hz~\cite{Yamamoto2023SAW,Jiang2023SAW}. Although these devices are included here for comparison, Fig.~\ref{fig:benchmarking} focuses on AlN- and AlScN-based FBARs, SMRs, and HBARs. Previously reported AlN FBARs achieve \(Qf\approx8.2~\mathrm{THz}\) at 285~K and \(20.7~\mathrm{THz}\) at 79~K~\cite{Zhao2024KuBandFBARLoss}. 

Cooling our epitaxial AlN FBAR from \(294\) to \(6.5~\mathrm{K}\) increases the de-embedded \(Q_{\max}\) from 245 to 1729. Operating near \(16~\mathrm{GHz}\), the device reaches a cryogenic \(Qf\) product of \(27.0~\mathrm{THz}\). These measurements are consistent with reduced temperature-dependent acoustic dissipation and extend the limited cryogenic data for fundamental-mode AlN FBARs above \(10~\mathrm{GHz}\).} Despite this improvement, the measured cryogenic \(Qf\) remains well below the calculated intrinsic Landau--Rumer limit of AlN at \(6.5~\mathrm{K}\). This gap suggests that extrinsic losses continue to limit the device and motivates further improvements in resonator geometry to increase \(Qf\). The epitaxial AlN platform also supports monolithic integration with GaN RF front-end electronics~\cite{Zhao2022XbandEpiBAWResonators}, providing a route to combining acoustic resonators and electronic components on the same platform.

\begin{figure}[h]
\centering
    \includegraphics[width=\textwidth]{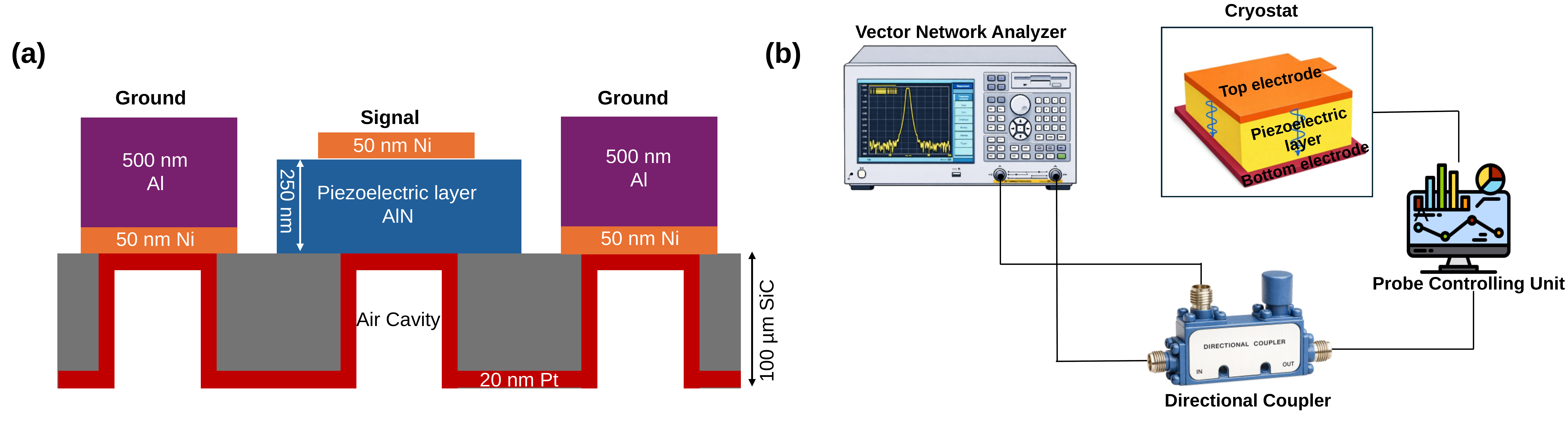}
\caption{(a) Cross-sectional schematic of the suspended AlN FBAR on SiC. (b) Cryogenic RF measurement setup used to measure $S_{11}$ and extract the unloaded maximum Bode quality factor, $Q_{\max}$, after open--short de-embedding.}
\label{fig:fbar_cryo_setup}
\end{figure}

\section{Device Architecture and Measurement}
Figure~\ref{fig:fbar_cryo_setup}(a) shows a cross-sectional schematic of the fabricated 15.6~GHz epitaxial AlN FBAR. {The fabrication procedure was adapted from the double-sided AlN-on-SiC process reported in Ref.~\cite{Zhao2024KuBandFBARLoss}. A 3-inch semi-insulating 4H--SiC wafer was mechanically thinned from \SI{350}{\micro\meter} to \SI{100}{\micro\meter}, followed by epitaxial AlN growth using molecular beam epitaxy. The AlN film was patterned by etching away the regions outside the resonator. A \SI{50}{\nano\meter} Ni top electrode was then deposited, followed by thicker Al probe pads. With the front surface protected, backside through-SiC etching exposed the underside of the AlN and formed an air cavity beneath the active region. After the protective layer was removed, Pt was deposited by thermal atomic layer deposition to conformally coat the exposed AlN and via sidewalls, forming a \SI{20}{\nano\meter} bottom electrode accessible through the ground contacts. The suspended Ni/AlN/Pt metal--insulator--metal (MIM) stack supports thickness-extensional vibration, while the air cavity reduces acoustic leakage into the substrate.} The remaining supports, or anchors, provide a pathway for mechanical energy leakage and therefore strongly influence the achievable $Q$ at a fixed resonance frequency.

Figure~\ref{fig:fbar_cryo_setup}(b) illustrates the cryogenic setup used to characterize the reflection response of the one-port FBAR. The device chip, approximately \(1\times1~\mathrm{cm}^2\), was mounted inside the cryostat and electrically contacted using an RF probe connected to the RF feedthroughs. { VNA Port~1 supplies the excitation through an external directional coupler to the FBAR. The coupler samples the wave reflected from the device and routes it to VNA Port~2 for detection. Thus, Port~2 acts as a receiver for the sampled reflected wave; both VNA connections serve the same device port. Measurements were performed on cooling from \SI{294}{\kelvin} to \SI{6.5}{\kelvin} and subsequent warming to \SI{294}{\kelvin}, with consistent results obtained in both directions.} The measured magnitude $M_{\mathrm{dB}}(f)$ (in dB) and phase $\Phi_{\mathrm{deg}}(f)$ (in degrees) are converted to linear magnitude and radians, respectively, and then combined to form $S_{11}(f)=|S_{11}(f)|e^{j\phi(f)}$.

\begin{figure}[h]
    \centering
    \includegraphics[width=\textwidth]{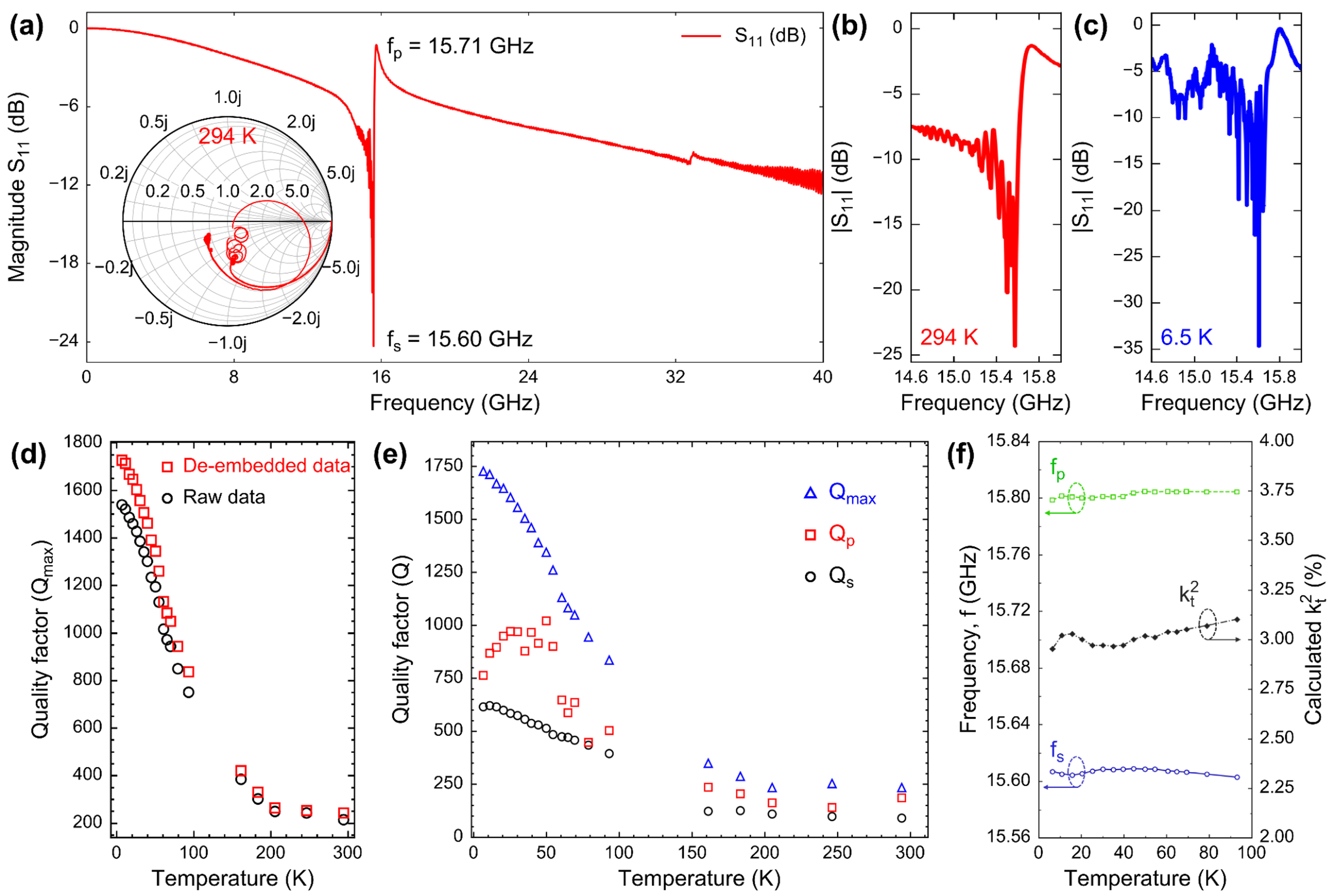}
\caption{Reflection spectra and temperature-dependent resonance characteristics of the 15.6 GHz AlN FBAR. Panels (a)--(c) show raw measurement data: (a) broadband magnitude of $S_{11}$ at 294 K, with the corresponding Smith chart shown in the inset; enlarged $S_{11}$ responses near resonance at (b) 294 K and (c) 6.5 K. (d) Temperature dependence of $Q_{\max}$ extracted from the raw measurements (black circles) and after de-embedding (red squares). (e) Temperature dependence of the de-embedded $Q_{\max}$ (blue triangles), $Q_{\mathrm{p}}$ (red squares), and $Q_{\mathrm{s}}$ (black circles), with $Q_{\mathrm{s}}$ and $Q_{\mathrm{p}}$ evaluated at the series and parallel resonance frequencies, respectively. (f) Temperature dependence of $f_{\mathrm{s}}$, $f_{\mathrm{p}}$, and the calculated electromechanical coupling coefficient $k_{\mathrm{t}}^{2}$.}
    \label{fig:Q_vs_temperature_figfinal}
\end{figure}

\section{Temperature-dependent Loss Limits in Microwave Acoustic Resonators}
Figure~\ref{fig:Q_vs_temperature_figfinal} presents the microwave reflection response and cryogenic characterization of the suspended epitaxial-AlN film bulk acoustic resonator (FBAR). Figure~\ref{fig:Q_vs_temperature_figfinal}(a) shows the measured small-signal RF reflection response of the fabricated AlN FBAR over the 0.05--40.05~GHz frequency range. A resonance is observed near 15.6~GHz in the reflection coefficient magnitude $S_{11}$, identifying the fundamental longitudinal thickness-extensional mode, whose resonance frequency is approximately $f_{0}\simeq v_{\mathrm{L}}/(2h)$, with additional corrections arising from electrode mass loading and the acoustic boundary conditions. The response exhibits a sharp minimum at the series resonance, $f_s=15.60$~GHz. A corresponding maximum is observed at the parallel resonance, $f_p=15.71$~GHz, where the impedance returns toward a high-value state. Figure~\ref{fig:Q_vs_temperature_figfinal}(a) inset represents Smith-chart for the 15.6~GHz AlN FBAR at 294~K in the complex plane through the normalized impedance $z=Z_{\mathrm{in}}/Z_{0}=(1+S_{11})/(1-S_{11})$. The resonance loop and rapid phase rotation indicate reversible energy exchange between the electrical field and the coherent acoustic mode. The resonance loop indicates excitation of the motional branch near the thickness-mode series resonance $f_s$, followed by a return toward high impedance near the antiresonance $f_p$. 

Figures~\ref{fig:Q_vs_temperature_figfinal}(b) and (c) show the reflection response ($|S_{11}|$) of the FBAR at $294~\mathrm{K}$ and $6.5~\mathrm{K}$, respectively. At $294~\mathrm{K}$, the principal series resonance occurs near $f_s\approx15.60~\mathrm{GHz}$, with a reflection minimum of approximately $-24~\mathrm{dB}$. Cooling to $6.5~\mathrm{K}$ shifts the main series and parallel resonances upward to approximately $f_s\approx15.61~\mathrm{GHz}$ and $f_p\approx15.80~\mathrm{GHz}$, consistent with cryogenic changes in the AlN elastic constants and acoustic velocity. {The ripples in the reflection spectra are attributed to nonuniform film thickness and uneven backside metallization, which perturb the acoustic boundary conditions and produce local variations in resonance frequency. These nonuniformities allow the primary thickness-extensional mode to couple to other longitudinal standing waves and closely spaced spurious modes.}

{Figure~\ref{fig:Q_vs_temperature_figfinal}(d) compares the temperature dependence of the maximum Bode quality factor, $Q_{\max}$, before and after de-embedding. The quality factor is calculated using the one-port Bode expression $Q_{\mathrm{Bode}}(f)=2\pi f\tau_{g}(f)|S_{11}(f)|/[1-|S_{11}(f)|^{2}]$, where $\tau_{g}(f)=-d[\arg S_{11}(f)]/d\omega$ is the reflection group delay~\cite{Feld2008NewFormulaForQ,Ruby2008ExtractingUnloadedQAcrossTech}. De-embedding increases $Q_{\max}$ from 216 to 245 at $294~\mathrm{K}$ and from 1540 to 1729 at $6.5~\mathrm{K}$. The de-embedded $Q_{\max}$ therefore improves by approximately 7.1 times upon cooling.

Figure~\ref{fig:Q_vs_temperature_figfinal}(e) compares $Q_{\max}$ with $Q_{\mathrm{s}}$ and $Q_{\mathrm{p}}$, evaluated at the series and parallel resonance frequencies, respectively. Whereas $Q_{\mathrm{s}}$ increases progressively on cooling, $Q_{\mathrm{p}}$ exhibits a non-monotonic temperature dependence, with a broad maximum at intermediate cryogenic temperatures.}

Figure~\ref{fig:Q_vs_temperature_figfinal}(f) shows the temperature dependence of the series and parallel resonance frequencies. For the fundamental thickness-extensional mode, their material-dependent contributions can be approximated as $f_s(T)\propto b(T)^{-1}\sqrt{c_{33}^{E}(T)/\rho(T)}$ and $f_p(T)\propto b(T)^{-1}\sqrt{c_{33}^{D}(T)/\rho(T)}$, where $b(T)$ and $\rho(T)$ are the AlN thickness and mass density, respectively. Here, $c_{33}^{E}$ and $c_{33}^{D}$ denote the longitudinal stiffness constants under constant-electric-field and constant-electric-displacement conditions, respectively, and are related through $c_{33}^{D}=c_{33}^{E}+e_{33}^{2}/\varepsilon_{33}^{S}$, where $e_{33}$ is the longitudinal piezoelectric coefficient and $\varepsilon_{33}^{S}$ is the dielectric permittivity at constant strain.

Cooling reduces the AlN thickness, shortening the acoustic propagation path and increasing both resonance frequencies. At the same time, the AlN density increases, reducing the acoustic velocity and shifting the frequencies downward. These smaller, opposing effects largely cancel, leaving the increase in AlN stiffness as the dominant contribution~\cite{GERLICH1986437,Zhao2024KuBandFBARLoss}.

Anharmonic lattice vibrations soften the elastic constants at elevated temperatures. Therefore, cooling increases $c_{33}^{E}$ and $c_{33}^{D}$ as the thermal-phonon population is suppressed. Accordingly, $c_{33}^{E}$ is expected to increase by only $\sim\SIrange{0.03}{0.05}{\giga\pascal}$ between \SI{100}{\kelvin} and \SI{1}{\kelvin}, where both $f_s$ and $f_p$ should approach saturation~\cite{Varshni1970Elastic,Zhao2024KuBandFBARLoss}. 

Thus, both $f_s$ and $f_p$ are expected to increase slightly as the temperature decreases and then saturate at low temperature. The measured frequencies follow the expected temperature dependence, while the small non-monotonic shift in $f_p$ likely arises from resonance-extraction uncertainty and mass loading effects. The effective electromechanical coupling coefficient, calculated as $k_t^2=\frac{\pi}{2}\frac{f_s}{f_p}\tan\!\left[\frac{\pi}{2}\frac{f_p-f_s}{f_p}\right]$, remains within approximately $2.95\%$--$3.10\%$ between 6.5 K and 100 K. It provides a measure of electrical-to-acoustic energy conversion, with a larger $f_p-f_s$ separation indicating stronger coupling.

\begin{figure}[h]
    \centering
    \includegraphics[width=\textwidth]{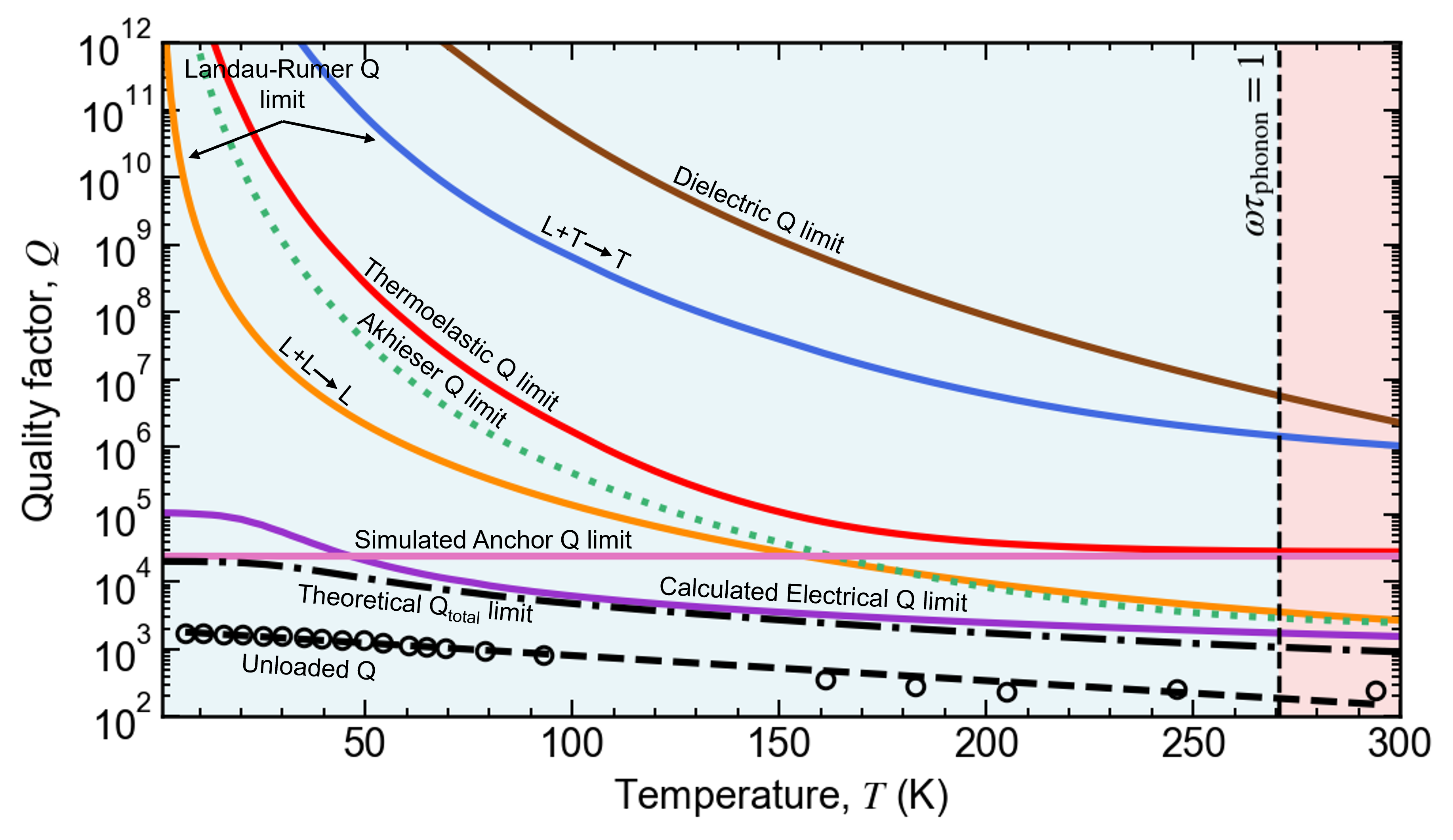}
\caption{Temperature dependence of the quality factor $Q$ of the \SI{15.6}{\giga\hertz} AlN FBAR. Open circles show the unloaded $Q_{\max}$ (the dashed line is a guide to the eye). The orange and blue curves represent the calculated Landau--Rumer limits associated with the $\mathrm{L}+\mathrm{L}\rightarrow\mathrm{L}$ and $\mathrm{L}+\mathrm{T}\rightarrow\mathrm{T}$ scattering channels, respectively. The green dotted, red, and brown curves show the Akhiezer, thermoelastic, and dielectric-loss limits, respectively. The magenta and purple curves show the simulated anchor-loss and electrical-loss limits, respectively. The black dash-dotted curve represents the total theoretical $Q$ limit obtained by combining the individual loss contributions.}
    \label{fig:Q_vs_temperature}
\end{figure}

The temperature-dependent quality-factor limits of acoustic resonators are summarized in Figures~\ref{fig:Q_vs_temperature}. The quality factor of an acoustic resonator is defined as~\cite{Capek2016ProcRSocA}
\begin{equation}
Q \;\equiv\; 2\pi\,
\frac{\text{Energy stored}}{\text{Energy dissipated per cycle}}
\label{eq:Q_def_cycle}
\end{equation}
We also deduced the quality factor ($Q$) in terms of power radiated, which was used in both the intrinsic and extrinsic loss terms. \(U_{stored}\) denotes the cycle-averaged stored energy and \(P_{\mathrm{loss}}\) the average dissipated power, defined as
\(P_{\mathrm{loss}}\equiv -dU_{stored}/dt\).
The energy dissipated over one oscillation period \(T_0\) is
\(\Delta U_{\text{cycle}}=\int_{t}^{t+T_0} P_{\mathrm{loss}}(t')\,dt'\).
For a high-\(Q\) resonator, \(P_{\mathrm{loss}}\) varies negligibly within a single cycle, so
\(\Delta U_{\text{cycle}}\approx P_{\mathrm{loss}}T_0\).
Using \(Q=2\pi\,U/\Delta U_{\text{cycle}}\) then gives
\(Q\approx 2\pi\,U/(P_{\mathrm{loss}}T_0)\), and substituting \(T_0=2\pi/\omega_0\), the Q is~\cite{Pierce1981Acoustics}
\begin{equation}
Q \;\equiv\;\ \omega_0 \frac{U_{stored}}{P_{\mathrm{loss}}}
\label{eq:Q_def}
\end{equation}
so that each loss mechanism can be interpreted as a channel that dissipates a fraction of the stored energy per oscillation. The combined theoretical limit plotted in figures~\ref{fig:Q_vs_temperature} is computed as~\cite{Yasumura2000QFactorsCantilevers,Ayazi2011EnergyDissipation} 
\begin{equation}
\begin{aligned}
\frac{1}{Q_{\mathrm{total}}(T)}
&=
\frac{1}{Q_{\mathrm{pp}}(T)}+
\frac{1}{Q_{\mathrm{TED}}(T)}+
\frac{1}{Q_{\mathrm{elec}}(T)}+
\frac{1}{Q_{\mathrm{diel}}(T)}+
\frac{1}{Q_{\mathrm{anchor}}}
\end{aligned}
\label{eq:Q_total_fracsum}
\end{equation}

To avoid double counting the two limits of phonon--phonon dissipation, the intrinsic material contribution is written using the single regime-appropriate quantity \(Q_{\mathrm{pp}}\). In an AlN FBAR, the intrinsic loss is governed by phonon–phonon and thermoelastic damping mechanisms, which require temperature-dependent thermophysical properties of AlN: volumetric heat capacity $C_v(T)$, thermal conductivity $\kappa(T)$, and linear thermal expansion coefficient $\alpha(T)$. In an insulating crystalline solid, the dominant contribution to the heat capacity arises from lattice vibrations (phonons). The Debye–Einstein~\cite{Sedmidubsky2009_CpCv_AIIIN} model captures lattice heat capacity over a broad temperature range. The corresponding volumetric heat capacity is given by
$C_v(T) = (\rho/M)\,C_{v,\mathrm{mol}}(T)$,
where $\rho$ is the mass density, $M$ is the molar mass, and $C_{v,\mathrm{mol}}(T)$ is the molar heat capacity. For AlN, $C_p(T)\approx C_v(T)$ up to $294~\si{K}$, since their difference is negligible in this temperature range~\cite{Wang2020AlN_ThermalExpansion_Cp}. The thermal conductivity $\kappa(T)$ is taken for AlN~\cite{Slack1987AlNThermalConductivity} and evaluated using a preserving cubic Hermite interpolation (PCHIP). We model the c-axis thermal expansion of AlN by fitting the measured c-lattice constant~\cite{Kroencke2008AlN_TEC_HRXRD} to a Debye form. The intrinsic quality factor depends on how these properties couple within each loss mechanism. 

The average lifetime of the thermal phonons is estimated by combining the thermal conductivity with the kinetic-theory relaxation-time approximation~\cite{Liu2005LossStudy,Maris1968UltrasonicAttenuationDirtyDielectrics} is
\begin{equation}
    \tau_{\mathrm{ph}}(T)
    =
    \frac{3\kappa(T)}{C_{v}(T)c_{D}^{2}}
    \label{eq:phonon_relaxation_time}
\end{equation}
The dimensionless product $\chi_{\mathrm{ph}}(T) \equiv \omega_{0}\tau_{\mathrm{ph}}(T)$ determines how the coherent acoustic period compares with the equilibration time of the thermal-phonon bath. The limits $\chi_{\mathrm{ph}} \gg 1$ and $\chi_{\mathrm{ph}} \ll 1$ correspond to the Landau--Rumer and Akhiezer regimes, respectively.

Landau--Rumer (LR) phonon-phonon scattering describes the high-frequency limit in which the coherent acoustic phonon oscillates faster than the thermal phonon population can re-equilibrate, i.e., \(\omega_{0}\tau_{\mathrm{ph}}\gg1\).  In this regime, the coherent thickness-extensional mode is attenuated through resolved three-phonon scattering events.  For a longitudinal acoustic mode, we have followed the model of Zhao \textit{et al.}~\cite{Zhao2024KuBandFBARLoss}:
\begin{align}
    L_{\mathrm{ac}}+L_{\mathrm{th}}&\rightarrow L_{\mathrm{th}},
    \label{eq:LR_LL_process}\\
    L_{\mathrm{ac}}+T_{\mathrm{th}}&\rightarrow T_{\mathrm{th}}
    \label{eq:LR_LT_process}
\end{align}
where the subscripts ``ac'' and ``th'' identify the coherent acoustic phonon and the thermal phonons, while \(L\) and \(T\) denote longitudinal and transverse directions. The attenuation coefficients implemented in the model are
\begin{gather}
    \alpha_{\mathrm{LL}}(T)
    =
    \left[
        \frac{8\pi^{5}\omega_{0}k_{B}^{4}T^{4}}
        {30\,\rho\,s_{L}^{6}h_{\mathrm{P}}^{3}}
    \right]
    \gamma_{\mathrm{LL}}^{2}
    \arctan\!\left(2\omega_{0}\tau_{\mathrm{ph}}\right),
    \label{eq:LR_LL_attenuation}\\[6pt]
    \alpha_{\mathrm{LT}}(T)
    =
    \left[
        \frac{8\pi^{5}\omega_{0}k_{B}^{4}T^{4}}
        {15\,\rho\,s_{L}^{2}s_{T}^{4}h_{\mathrm{P}}^{3}}
    \right]
    \gamma_{\mathrm{LT}}^{2}
    \arctan\!\left[
        \frac{
            2\omega_{0}\tau_{\mathrm{ph}}
            \left(s_{T}/s_{L}\right)
        }{
            1+
            \left(\omega_{0}\tau_{\mathrm{ph}}\right)^{2}
            \left[1-\left(s_{T}/s_{L}\right)^{2}\right]
        }
    \right]
    \label{eq:LR_LT_attenuation}
\end{gather}
where \(k_{B}\) is Boltzmann's constant, \(h_{\mathrm{P}}\) is Planck's constant, and \(\gamma_{\mathrm{LL}}\) and \(\gamma_{\mathrm{LT}}\) are effective mode-dependent Gr\"uneisen parameters. The Gr\"uneisen parameters quantify how strongly the frequencies of the thermal phonons are modulated by strain; because wurtzite AlN is elastically and anharmonically anisotropic, they should be understood as effective parameters for the propagation direction and polarization of the measured mode. The total LR attenuation and the associated quality-factor limit are
\begin{equation}
\begin{aligned}
Q_{\mathrm{LL}}(T)
    &= \frac{\omega_0}
             {2s_L\alpha_{\mathrm{LL}}(T)},
    \qquad
Q_{\mathrm{LT}}(T)
    &= \frac{\omega_0}
             {2s_L\alpha_{\mathrm{LT}}(T)}
\end{aligned}
\label{eq:LR_quality_factor}
\end{equation}
Equations~\eqref{eq:LR_LL_attenuation}--\eqref{eq:LR_quality_factor} explains the main LR trends.  The explicit \(T^{4}\) factor causes the phonon-scattering loss to collapse rapidly on cooling.  In the deep LR limit, the arctangent in the LL term approaches \(\pi/2\), so that \(\alpha_{\mathrm{LL}}\propto\omega_{0}T^{4}\) and \(Q_{\mathrm{LL}}\propto T^{-4}\) when the material parameters are treated as slowly varying.  

Akhiezer phonon-phonon scattering occurs in the opposite limit, $\omega_{0}\tau_{\mathrm{ph}}\ll1$. The strain of the coherent acoustic mode shifts the frequencies of the thermal phonons because of lattice anharmonicity, pushing their energy distribution away from equilibrium~\cite{Ghatge2018Anharmonicity}. Frequent phonon--phonon collisions restore equilibrium, producing entropy and converting part of the coherent-mode energy into heat. The Akhiezer expression used here is~\cite{Hwang2010LOWTQ}
\begin{equation}
    Q_{\mathrm{AKE}}(T)
    =
    \left(
        \frac{2\pi\rho c_{D}^{2}}
        {\gamma_{\mathrm{LL}}^{2}}
    \right)
    \frac{
        1+\left[\omega_{0}\tau_{\mathrm{ph}}(T)\right]^{2}
    }{
        \omega_{0}\tau_{\mathrm{ph}}(T)\,T\,C_{v}(T)
    }
    \label{eq:Akhiezer_quality_factor}
\end{equation}

For the present AlN parameters, however, \(T C_{v}(T)\) increases and \(\tau_{\mathrm{ph}}(T)\) decreases as room temperature is approached, moving the device toward the maximum of the relaxation function and increasing the calculated loss. At \(f_{0}=\SI{15.6}{\giga\hertz}\), the calculated regime parameter decreases from approximately \(10^{4}\) at \(T=\SI{6.5}{\kelvin}\) to \(\omega_{0}\tau_{\mathrm{ph}}=1\) near \(\SI{270}{\kelvin}\) to \(0.73\) at \(T=\SI{294}{\kelvin}\). At cryogenic temperatures, the resonator therefore operates in the LR regime. At room temperature, the FBAR lies on the Akhiezer side of the limit.  Landau--Rumer and Akhiezer phonon--phonon scattering are not two independent terms in the inverse-\(Q\) sum.  We therefore denote the applicable phonon--phonon limit by \(Q_{\mathrm{pp}}\):
\begin{equation}
    Q_{\mathrm{pp}}^{-1}(T)
    \simeq
    \begin{cases}
        Q_{\mathrm{LR}}^{-1}(T),
        & \omega_{0}\tau_{\mathrm{ph}}\gg1,\\[3pt]
        Q_{\mathrm{AKE}}^{-1}(T),
        & \omega_{0}\tau_{\mathrm{ph}}\ll1
    \end{cases}
    \label{eq:phonon_phonon_regime_selection}
\end{equation}

The relaxation time used in thermoelastic damping has a different origin.  It is the time required for heat to diffuse over a characteristic distance \(b\), which is taken to be the AlN thickness for the one-dimensional FBAR estimate:
\begin{equation}
    \tau_{\mathrm{TED}}(T)
    =
    \left(\frac{b}{\pi}\right)^{2}
    \frac{C_{p}(T)}{\kappa(T)}
    \label{eq:TED_relaxation_time}
\end{equation}

Thermoelastic damping (TED) is generated by spatial heat flow rather than by the direct equilibration of the thermal-phonon distribution.  Compression locally raises the equilibrium temperature, while tension lowers it.  If the acoustic mode creates neighboring regions with different strain, it also creates a periodic temperature gradient.  Heat flows from the compressed regions toward the tensile regions, but thermal diffusion is not instantaneous.  The resulting phase lag between stress and strain converts part of the coherent elastic energy into heat during every cycle~\cite{Zener1938ThermoelasticInternalFriction}. In the 1D Zener approximation, the strength of the thermoelastic coupling is
\begin{equation}
    \Delta_{\mathrm{TED}}(T)
    =
    \frac{
        E_{\mathrm{eff}}(T)\alpha_{\mathrm{th}}^{2}(T)T
    }{
        C_{p}(T)
    }
    \label{eq:TED_relaxation_strength}
\end{equation}
where \(E_{\mathrm{eff}}\) is the elastic modulus appropriate to the strain
component of the resonant mode.  Combining
Eq.~\eqref{eq:TED_relaxation_strength} with the diffusion time in
Eq.~\eqref{eq:TED_relaxation_time} gives~\cite{Liu2005LossStudy,Rodriguez2019}
\begin{align}
    Q_{\mathrm{TED}}(T)
    &=
    \frac{C_{p}(T)}
    {E_{\mathrm{eff}}(T)\alpha_{\mathrm{th}}^{2}(T)T}
    \frac{
        1+\left[\omega_{0}\tau_{\mathrm{TED}}(T)\right]^{2}
    }{
        \omega_{0}\tau_{\mathrm{TED}}(T)
    }
    \label{eq:TED_quality_factor}
\end{align}

For a thickness-extensional FBAR mode, Eqs.~\eqref{eq:TED_relaxation_time} and \eqref{eq:TED_quality_factor} provide a first-order estimate. The exact TED depends on the three-dimensional strain field, crystallographic anisotropy, and the actual distance between regions of opposite thermoelastic temperature perturbation~\cite{Dabas2026Akhiezer}. 

An alternating electric field also dissipates energy in the AlN layer, resulting in dielectric loss. Wurtzite AlN is a polar, noncentrosymmetric crystal with space group \(P6_{3}mc\).  The electric field can therefore couple to the acoustic-phonon system, producing an intrinsic lattice absorption even in a chemically perfect, leakage-free crystal. Following the framework of Gurevich and Tagantsev, the low-temperature intrinsic loss of a noncentrosymmetric dielectric contains a quasi-Debye contribution and a three-quantum contribution~\cite{Gurevich1991IntrinsicDielectricLoss}. For the field component parallel to the AlN \(c\)-axis, we write the model in the explicit form
\begin{gather}
    \tau_{d}(T)
    =
    \left[
        \left(
            \frac{k_{B}T}
            {M_{\mathrm{atom}}v_{D}^{2}}
        \right)
        \left(
            \frac{T}{\Theta_{D}}
        \right)^{3}
        \left(
            \frac{k_{B}T}{\hbar}
        \right)
    \right]^{-1}
    \label{eq:dielectric_relaxation_time}
    \\[4pt]
    \tan\delta_{\mathrm{QD},33}(T)
    =
    \frac{1}{\varepsilon_{33}^{\prime}}
    \frac{(k_{B}T)^{4}}
    {\rho v_{D}^{5}\hbar^{3}}
    \frac{\omega_{0}\tau_{d}(T)}
    {1+\left[\omega_{0}\tau_{d}(T)\right]^{2}}
    \label{eq:quasi_Debye_dielectric_loss}
    \\[4pt]
    \tan\delta_{3\mathrm{q},33}(T)
    =
    \frac{1}{\varepsilon_{33}^{\prime}}
    \frac{\omega_{0}^{3}k_{B}T}
    {\rho v_{D}^{5}}
    \label{eq:three_quantum_dielectric_loss}
    \\[4pt]
    \tan\delta_{33,\mathrm{int}}(T)
    =
    \tan\delta_{\mathrm{QD},33}(T)
    +
    \tan\delta_{3\mathrm{q},33}(T)
    \label{eq:intrinsic_dielectric_loss}
\end{gather}
Here, \(\varepsilon_{33}^{\prime}\) is the real relative-permittivity component, \(v_{D}\) is the acoustic velocity used in the dielectric-loss approximation, \(\Theta_{D}\) is the Debye temperature, \(M_{\mathrm{atom}}\) is the average mass per atom, and \(\tau_{d}\) is the thermal-phonon relaxation time used in the dielectric model. 

The quasi-Debye term is a relaxation process. The three-quantum term represents direct absorption involving the AC field quantum and two lattice phonons. In the simplified expression above, it grows as \(\omega_{0}^{3}T\).  Their sum establishes an ideal intrinsic lattice limit for \(\tan\delta\). The approximation used in understanding dielectric loss is \(Q_{\mathrm{diel}}(T)=1/\tan\delta_{33,\mathrm{int}}(T)\). Real films generally exhibit a measured loss tangent that is larger than the ideal intrinsic lattice value because of dislocations, point defects, grain or domain boundaries, interfaces, surface contamination, and leakage conduction~\cite{Alford2001DielectricLossOxides}. Equivalently, such extrinsic dielectric processes reduce the measured \(Q_{\mathrm{diel}}\).

The RF current used to charge and discharge the piezoelectric transducer produces Joule heating in the finite-resistance electrodes, resulting in electrical loss. Near series resonance, the electrical branch may be represented by a motional capacitance \(C_{m}\) and an effective series resistance \(R_{s}\).  The resulting electrical-loss-limited quality factor is~\cite{Zou2022}
\begin{equation}
    Q_{\mathrm{elec}}(T)
    =
    \frac{1}
    {\omega_{0}C_{m}R_{s}(T)}
    \label{eq:electrical_loss_Q}
\end{equation}
The Ni top electrode and Pt bottom electrode are treated separately. If their resistances are in series with the same motional current,
\begin{align}
    \frac{1}{Q_{\mathrm{elec}}(T)}
    &=
    \frac{1}{Q_{\mathrm{top}}(T)}
    +
    \frac{1}{Q_{\mathrm{bot}}(T)}
    \label{eq:electrode_inverse_Q_sum}\\
    Q_{j}(T)
    &=
    \frac{1}{\omega_{0}C_{m}R_{j}(T)}
    \label{eq:individual_electrode_Q}
\end{align}
For radial current flow through a circular metal film, the first-order resistance is
\begin{equation}
    R_{\mathrm{spread},j}(T)
    =
    \frac{\varrho_{j}(T)}{2\pi t_{j}}
    \ln\!\left(\frac{a_{j}}{r_{c,j}}\right),
    \label{eq:circular_spreading_resistance}
\end{equation}
where \(\varrho_{j}\) is the electrical resistivity, \(t_{j}\) is the metal thickness, \(a_{j}\) is the outer current-spreading radius, and \(r_{c,j}\) is the effective contact radius. The Pt resistivity is obtained by interpolating the cryogenic data compiled in NBS Technical Note 365~\cite{Hall1968_NBSTN365}.  Over the fitted temperature interval, the Ni resistivity is represented by
\begin{equation}
    \varrho_{\mathrm{Ni}}(T)
    =
    \varrho_{\mathrm{res}}
    +a_{2}T^{2}
    +a_{4}T^{4},
    \label{eq:Ni_resistivity_polynomial}
\end{equation}
where \(\varrho_{\mathrm{res}}\) is the residual resistivity and the polynomial coefficients approximate the relevant portion of the Bloch--Gr\"uneisen temperature dependence~\cite{Schwerer1968NiResistivity}. The residual resistance of a nanometer-scale film, however, can differ from the bulk value. 

\begin{figure}[h]
    \centering
    \includegraphics[width=\textwidth]{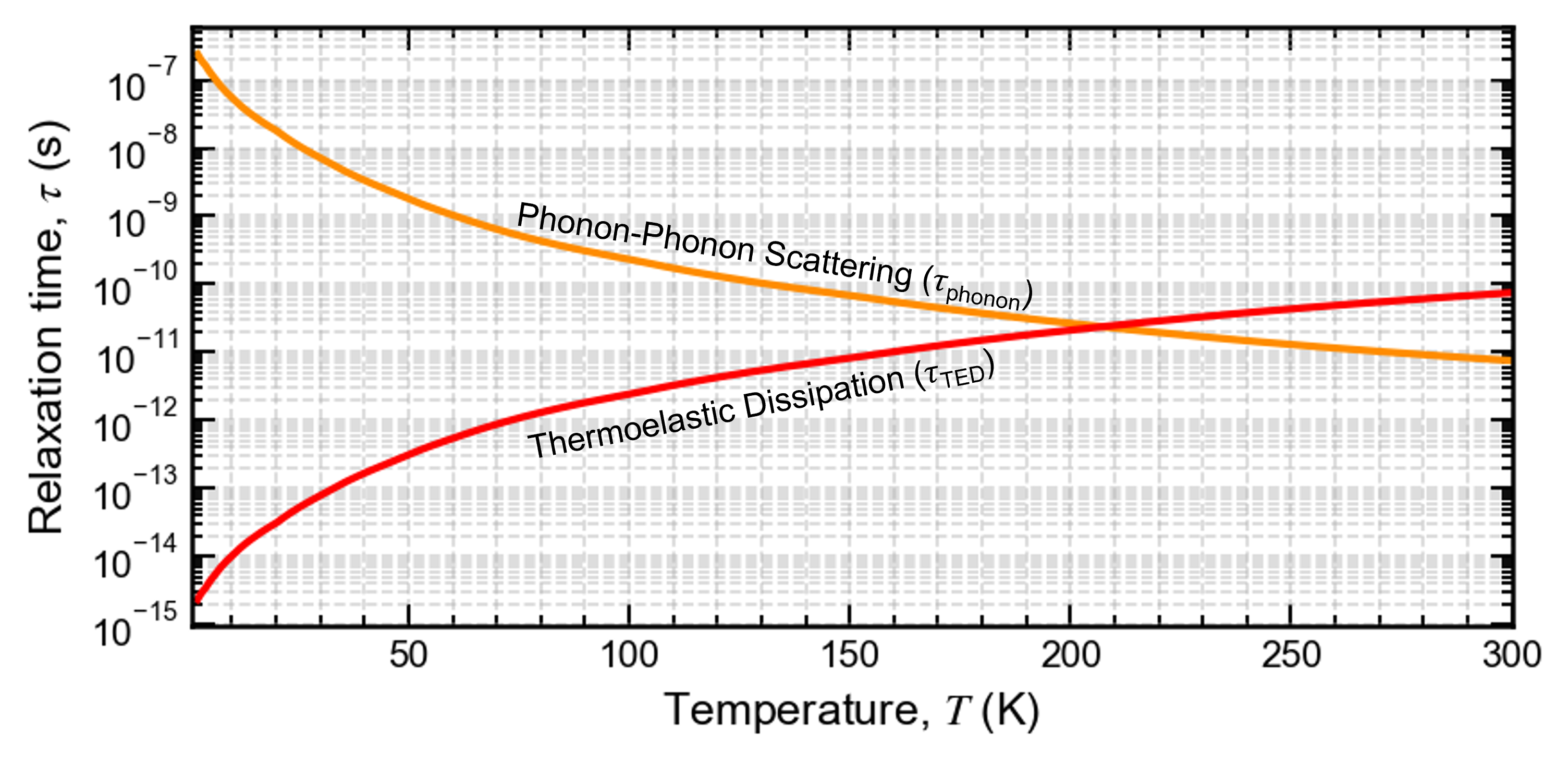}
\caption{Temperature dependence of the relaxation times associated with phonon--phonon scattering and thermoelastic dissipation. The phonon relaxation time is modeled as $\tau_{\mathrm{phonon}}=3\,\kappa(T)/\!\big(C_{v}(T)\,c_{D}^{2}\big)$, while the thermoelastic relaxation time is $\tau_{\mathrm{TED}}=\left(b/\pi\right)^{2} C_{p}(T)/\kappa(T)$. Here, $b$ is the film thickness, $\kappa(T)$ is the thermal conductivity~\cite{Slack1987AlNThermalConductivity}, $C_v(T)$ is the volumetric heat capacity~\cite{Sedmidubsky2009_CpCv_AIIIN}, and $C_p(T)$ is the constant-pressure heat capacity.}
    \label{fig:tau_vs_T}
\end{figure}

Anchor loss arises when acoustic energy escapes from the vibrating resonator through its supports and propagates into the substrate as elastic waves. Its magnitude depends on the mode shape, the support geometry, and the mechanical response of the substrate. Following the energy-balance approach used in attachment-loss models~\cite{Judge2007AttachmentLoss}, the corresponding quality factor is $Q_{\mathrm{anchor}}=\omega_0 U_{\mathrm{stored}}/P_{\mathrm{anchor}}$, where $\omega_0=2\pi f_0$, $U_{\mathrm{stored}}$ is the stored modal energy, and $P_{\mathrm{anchor}}$ is the time-averaged power carried away through the supports.

For a longitudinal-mode estimate, the stored mechanical energy is approximated as
$U_{\mathrm{stored}}\simeq \frac{1}{2}\rho A_{\mathrm{piezo}}t_{\mathrm{AlN}}\omega_0^2u_0^2$, where $\rho$ is the AlN density, $A_{\mathrm{piezo}}$ is the effective area over which the vibrational energy is stored, $t_{\mathrm{AlN}}$ is the AlN thickness, and $u_0$ is an effective peak displacement amplitude. The motion transmitted to the support generally differs from
that in the active region. We describe this difference using
$u_{\mathrm{sup}}=\gamma_{\mathrm{sup}}u_0$, where
$u_{\mathrm{sup}}$ is the effective peak displacement amplitude
associated with radiation at the support. Approximating the
support strain as $u_{\mathrm{sup}}/l_{\mathrm{sup}}$ gives
$\sigma_{\mathrm{sup}}\simeq
(c_{33}/l_{\mathrm{sup}})\gamma_{\mathrm{sup}}u_0$,
where $c_{33}$ is the effective longitudinal elastic stiffness
used in this approximation and $l_{\mathrm{sup}}$ is the
effective length over which the support displacement produces
strain.

If the support is further approximated as a longitudinal
plane-wave radiation port, the radiated power is
$P_{\mathrm{anchor}}\simeq
[A_{\mathrm{sup,tot}}/(2Z_{\mathrm{SiC}})]
\sigma_{\mathrm{sup}}^2
=
[A_{\mathrm{sup,tot}}/(2Z_{\mathrm{SiC}})]
[(c_{33}/l_{\mathrm{sup}})\gamma_{\mathrm{sup}}u_0]^2$,
where $Z_{\mathrm{SiC}}$ is the longitudinal acoustic impedance
of SiC and $A_{\mathrm{sup,tot}}$ is the total radiating support
area. For $N$ equivalent support regions,
$A_{\mathrm{sup,tot}}=N A_{\mathrm{sup,1}}$.
The factor of $1/2$ accounts for time averaging when the stress
is expressed as a peak amplitude. This approximation treats
the radiation as longitudinal and neglects the detailed
spatial variation of stress and velocity across the support.

Substituting the stored energy and radiated power into the
energy-balance expression eliminates the arbitrary displacement
amplitude and gives
$Q_{\mathrm{anchor}}\simeq
\rho A_{\mathrm{piezo}}t_{\mathrm{AlN}}
\omega_0^3Z_{\mathrm{SiC}}l_{\mathrm{sup}}^2/
(A_{\mathrm{sup,tot}}c_{33}^2\gamma_{\mathrm{sup}}^2)$.
For the annular support approximation used here,
$A_{\mathrm{piezo}}=\pi R_o^2$ and
$A_{\mathrm{sup,tot}}=\pi(R_o^2-R_i^2)$,
where $R_o$ is the outer radius of the AlN region and $R_i$
is the inner radius of the supporting annulus. The resulting
expression is
\begin{equation}
    Q_{\mathrm{anchor}}
    \simeq
    \frac{
        \rho\,R_o^2 t_{\mathrm{AlN}}
        \omega_0^3 Z_{\mathrm{SiC}}l_{\mathrm{sup}}^2
    }{
        \left(R_o^2-R_i^2\right)
        c_{33}^2\gamma_{\mathrm{sup}}^2
    }.
    \label{eq:anchor_quality_factor_annular}
\end{equation}

These expressions explain how support motion contributes to anchor radiation. In particular, stronger motion transmitted to the support increases the radiated power and lowers $Q_{\mathrm{anchor}}$. However, $\gamma_{\mathrm{sup}}$ depends on the displacement distribution of the selected mode and on how the active-region and support amplitudes are evaluated. The effective strain length $l_{\mathrm{sup}}$ also depends on the deformation near the support. These quantities are therefore difficult to determine reliably from the device dimensions alone. We retain the analytical expressions to describe the underlying physics. {To understand those parameters physically and estimate anchor radiation, we used a finite-element eigenfrequency model in COMSOL Multiphysics. The circular resonator was represented by a two-dimensional axisymmetric model, with Solid Mechanics and Electrostatics coupled through piezoelectricity. The model included the AlN, Ni, and Pt electrodes, and the supporting SiC. Perfectly matched layers (PML) were placed along the lower and outer edges of the retained substrate to represent outgoing elastic waves. Mechanical damping, dielectric loss, thermoelastic coupling, and resistive electrical losses were excluded so that the calculation isolated radiation through the support. The finite-element solution determines the support motion and its coupling to the substrate.

For the selected thickness-extensional mode, the stored energy decays as $U(t)=U_0e^{-2|\omega_i|t}$, where $\widetilde{\omega}=\omega_r+i\omega_i$ is the complex angular eigenfrequency. When anchor radiation is the only loss mechanism, the radiated power is $P_{\mathrm{anchor}}=-dU/dt=2|\omega_i|U$. Substituting this into $Q_{\mathrm{anchor}}=\omega_r U/P_{\mathrm{anchor}}$ and using $\widetilde{\omega}=2\pi\widetilde f$ gives the simulated anchor-loss quality factor~\cite{Steeneken2007SupportLoss},
\begin{equation}
    Q_{\mathrm{anchor}}^{\mathrm{sim}}
    =
    \frac{|\operatorname{Re}(\widetilde{\omega})|}
         {2|\operatorname{Im}(\widetilde{\omega})|}
    =
    \frac{|\operatorname{Re}(\widetilde f)|}
         {2|\operatorname{Im}(\widetilde f)|}.
    \label{eq:anchor_quality_factor_sim}
\end{equation}

The real part gives the oscillation frequency, while the imaginary part describes the decay of the modal amplitude. The eigenfrequency search was centered at $15.23~\mathrm{GHz}$, and the selected mode yielded $\widetilde f\simeq(15.23+3.06\times10^{-4}i)~\mathrm{GHz}$. The corresponding estimate is
\begin{equation}
    Q_{\mathrm{anchor}}^{\mathrm{sim}}
    =
    \frac{1.523\times10^{10}}
         {2(3.06\times10^{5})}
    \simeq 2.49\times10^4.
    \label{eq:anchor_quality_factor_sim_value}
\end{equation}
This simulated estimate is used for the anchor-loss contribution in the $Q$ versus $T$ model.}

{Figure~\ref{fig:Q_vs_temperature} compares the unloaded $Q_{\max}$ with the calculated limits from individual loss mechanisms. The black dash-dotted curve represents the total theoretical limit, obtained by adding the inverse quality factors of the included loss mechanisms. At the lowest temperatures, anchor loss sets the upper limit within the model, while electrical loss becomes increasingly important as the temperature rises. Cooling strongly reduces the calculated intrinsic phonon loss and increases the measured $Q_{\max}$. However, de-embedded values remain below the total theoretical curve. Several device-level contributions may explain this remaining difference. De-embedding reduces the pad and interconnect parasitics, but the TSV structure beneath the AlN remains part of the device, and its electrical contribution may not be fully omitted. The electrical-loss estimate also depends on the bulk Ni and Pt resistivities. In the thin-film electrodes, surface and grain-boundary scattering, defects, and deposition conditions can change the resistivity~\cite{Kim2019PtElectrical,deVries1987NiResistivity}. Their residual resistance can be higher even at cryogenic temperatures. Nonuniform film thickness and coupling to spurious modes may further reduce the measured $Q$.

Near room temperature, electrical loss sets the upper bound on $Q$, indicating an increase in electrode series resistance. This is consistent with the lower $Q_{\mathrm{s}}$ at higher temperatures in Figure~\ref{fig:Q_vs_temperature_figfinal}(e), since series resistance reduces the quality factor at series resonance. The increase in $Q_{\mathrm{s}}$ upon cooling is higher with reduced resistive loss. Together, these results show that the measured performance reflects losses in the electrodes and surrounding device structure in addition to intrinsic dissipation in AlN.}

Figure~\ref{fig:tau_vs_T} helps distinguish thermoelastic dissipation (TED) from phonon--phonon scattering in the loss comparison shown in Figure~\ref{fig:Q_vs_temperature}. Although both involve interactions among strain, temperature, and phonons, they arise from different physical processes. For TED, $\tau_{\mathrm{TED}}$ is a diffusion time associated with heat flow over the thickness of the piezoelectric material. In this case, loss arises because strain and thermal expansion create spatially non-uniform equilibrium temperatures, and the finite time required for heat to diffuse between compressed and tensile regions produces a phase lag between stress and strain. As a result, TED is strongly geometry-dependent and is maximized near $\omega_0\tau_{\mathrm{TED}}\sim 1$. By contrast, $\tau_{\mathrm{phonon}}$ in the Akhiezer and Landau--Rumer models is the characteristic time for thermal phonons to exchange and redistribute energy through phonon--phonon collisions. It represents the average time between collisions of thermal phonons inside the crystal, or equivalently, how quickly the disturbed ``phonon gas'' relaxes back to equilibrium through phonon--phonon scattering. Therefore, $\tau_{\mathrm{phonon}}$ is a microscopic, material-controlled relaxation time, whereas $\tau_{\mathrm{TED}}$ is a macroscopic, geometry-dependent diffusion time.

\section{Conclusion}
Here, we fabricate an epitaxial AlN FBAR and use it to identify the dominant acoustic loss mechanisms at cryogenic temperatures and frequencies above \SI{10}{\giga\hertz}, where experimental data remain scarce. Small-signal measurements of its \SI{15.6}{\giga\hertz} fundamental thickness-extensional mode show that the maximum unloaded quality factor increases from $Q_{\max}=245$ to $Q_{\max}=1729$ ($Qf\approx\SI{27}{\tera\hertz}$) at \SI{6.5}{\kelvin}. A physics-based model that includes phonon--phonon dissipation, thermoelastic damping, dielectric loss, electrode resistance, and anchor loss helps explain the measured $Q(T)$ trend. {Finite-element eigenfrequency analysis gives $Q_{\mathrm{anchor}}^{\mathrm{sim}}\simeq2.49\times10^4$, indicating that anchor loss sets an upper bound at cryogenic temperatures while electrical loss limits $Q$ at higher temperatures.} The dimensionless phonon-regime parameter, $\chi_{\mathrm{ph}}=\omega_{0}\tau_{\mathrm{phonon}}$, decreases from approximately $10^{4}$ at \SI{6.5}{\kelvin} to $0.73$ at \SI{294}{\kelvin}, crossing unity near \SI{270}{\kelvin}. The FBAR therefore operates in the Landau--Rumer regime at cryogenic temperatures and enters the Akhiezer regime near room temperature.

The reported $Q$ values remain below the model-predicted upper bounds. The loss terms in this model are evaluated in the small-signal linear regime. Power-dependent effects, including self-heating, nonlinear acoustic response, and two-level-system saturation, are not included. The relaxation times $\tau_{\mathrm{phonon}}$ and $\tau_{\mathrm{TED}}$ describe different physical processes: $\tau_{\mathrm{phonon}}$ represents microscopic phonon collisions, while $\tau_{\mathrm{TED}}$ is set by geometry-dependent thermal diffusion. Together, the measurements and model separate intrinsic material losses from device-level losses and provide realistic bounds for $Q(T)$ in high-frequency FBARs. {Lower-resistance electrodes, improved acoustic isolation, and more uniform suspended films provide routes toward higher-$Q$ epitaxial AlN FBARs for upper-mid-band RF front ends and cryogenic microwave circuits.}

\ack
W.Z. and H. G. were supported by W.Z.’s start-up fund from Rensselaer Polytechnic Institute. Part of the work of W.Z. was supported in part by the Semiconductor Research Corporation (SRC) and the Defense Advanced Research Projects Agency (DARPA) through the Joint University Microelectronics Program (JUMP), carried out in the Cornell NanoScale Facility, a National Nanotechnology Coordinated Infrastructure (NNCI) member partly funded by the National Science Foundation (NSF) under Grant NNCI-2025233. 

\section*{Author contributions}

\noindent
\textbf{Hemant Gulupalli:} Conceptualization (lead); Methodology (equal);
Investigation (lead); Formal analysis (lead); Visualization (lead);
Writing -- original draft (lead); Writing -- review \& editing (equal).
\textbf{Navnil Choudhury:} Investigation (supporting); Data curation
(supporting); Validation (supporting); Writing -- review \& editing
(supporting).
\textbf{Jiacheng Xie:} Investigation (supporting).
\textbf{Yufeng Wu:} Investigation (supporting).
\textbf{Huili Grace Xing:} Funding acquisition (equal); Supervision (equal).
\textbf{Hong X. Tang:} Resources (equal).
\textbf{Debdeep Jena:} Conceptualization (equal); Funding acquisition (equal);
Supervision (equal).
\textbf{Kanad Basu:} Funding acquisition (equal); Supervision (equal).
\textbf{Wenwen Zhao:} Conceptualization (lead); Methodology (equal);
Funding acquisition (equal); Supervision (lead);
Writing -- review \& editing (equal).
\data{The data and code that support the findings of this study are available from the authors on reasonable request.}

\bibliographystyle{iopart-num-long}
\bibliography{References}

\end{document}